\documentstyle[12pt]{article}
\newcommand{\beq}{\begin{equation}}
 \newcommand{\eq}{\end{equation}}
\begin{document}
\pagenumbering{roman}
\begin{flushright}
JINR Communication \\ E2--96--15 \\ February 1996
\end{flushright}

\vspace{4cm}
\begin{center}
{\large\bf THE BOGOLIUBOV RENORMALIZATION GROUP}

 \vspace{3mm}
Second English printing  \\

 \vspace{10mm}
D.V.~Shirkov \\
\vspace{5mm}

     N.N.Bogoliubov Laboratory of Theoretical Physics, \\ JINR, Dubna \\
\vspace{5mm}
\end{center}

 \vspace{5mm}

\abstract{ We begin with personal notes describing the atmosphere of
"Bogoliubov renormalization group" birth. Then we expose the history
of RG discovery in the QFT and of the RG method devising in the
mid-fifties. The third part is devoted to proliferation of RG ideas
into diverse parts of theoretical physics.
We conclude with discussing the perspective of RG method further
development and its application in mathematical physics.} %

\newpage

\centerline{\small\sc Comment to the second English printing}


{\scriptsize
    This article originally appeared in August 1994 in Russian as a
preprint of Joint Institute for Nuclear Research No. P2-94-310
being simultaneously submitted to 'Uspekhi Math. Nauk'. Unfortunately,
in the second part of published version ['Успехи Мат. Наук', т.49,
No.5, (1994) 147-164] by some pure technical reasons there appeared
many (more than 25)  misprints related to the references of papers.
In spite of the author's signal for the UMN Editorial Board, these
errors have been reproduced in the American translation.}

{\scriptsize
       	Moreover, this latter English-language publication ['Russian
Math. Surveys', 49:5 (1994) 155-176], due to the translator's poor
qualification both in subject terminology and Russian, contains a lot
(around 50) of errors distorting the author's text.}

{\scriptsize
       By these reasons the author decided to present corrected
English text. It results from the editing of the RMS
publication and is adequate to the Russian preprint P2-94-310. }


\tableofcontents

\newpage
\pagenumbering{arabic}

\section{I N T R O D U C T I O N} 

\subsection{Quantum fields}
       Nikolai Nikolaevich Bogoliubov (N.N. in what follows) took up
problems in quantum field theory (QFT) in real earnest in the late 40s,
possibly under the influence of the well--known series of articles by
Schwinger which were presented at N.N.'s seminar in the Steklov
(Mathematical) Institute. In any case, the first publications on
quantum field theory
by N.N., which appeared in the early 50s\cite{nn-50/51}, were devoted to
variational--derivative equations of Tomonaga--Schwinger type and
were based on the axiomatic definition of the scattering matrix
as a functional of the interaction domain $g(x)$, generalizing
Schwinger's surface function $\sigma(t)$. \par
       In the first half of the 50s N.N. made an active entry from
the mathematical direction into a developing science, namely,
renormalizable QFT, progressing more rapidly and reaching more
deeply than other scientists who moved from mathematics into
theoretical physics. As is known, he developed his own renormalization
method based on the theory of Sobolev--Schwartz distributions. His
approach makes it possible to dispense with ``bare" fields and
particles and the physically unsatisfactory picture of infinite
renormalizations. \par
       It was N.N.'s custom from time to time  to present lectures
containing surveys of large portions of QFT such as
``renormalizations'', ``functional integral'', or ``surface divergences''.
Those who listened to the whole series of surveys were under the
impression that N.N. ``saw'' these outwardly so different
fragments from a single viewpoint, perceiving them as parts of the
same picture. \par
     This was at a time when the pre-war edition of Heitler's
``Quantum Theory of Radiation'' served as a textbook on the theory
of elementary particles. Akhiezer and Berestetskii's ``Quantum
Electrodynamics'' (1953) and the first volume of ``Mesons and
Fields'' by Bethe, Hoffman and Schweber (1955) were yet to appear. \par
   One day in the autumn of 1953, being under the influence of one
of these lectures, I asked: ``N.N., why don't you write a textbook on
the new QFT?''. His answer was: ``That's not a bad idea, perhaps we
should do it together?''. At first I did not take it seriously. It
should be explained that it was only in May of that memorable year
that one of the co--authors of the future book defended his candidate
dissertation in diffusion and neutron thermalization theory and
he did not have a single publication in quantum field theory, while
already in October of the same year the other became an academician.
\par   Nevertheless, the conversation was resumed the following
week and we begun to discuss the details of the project. The time
frame of these events is quite well defined, firstly, by the fact
that the above conversation took place in a car when going to N.N.'s
flat in Shchukinskii passage (near the Kurchatov Institute),
that is, before N.N. moved to the Moscow State University high--rise
building in Lenin Hills at the end of 1953. Secondly, Akhiezer and
Berestetskii's book had appeared just before our proposal was
submitted at Gostekhizdat at the beginning of 1954. At the same time,
the first version of the subsequent presentation of the axiomatic
scattering $S$-matrix was put forward for publication in Uspekhi Fiz.
Nauk at the end of 1954. \par
     The initial draft of the book, apart from an introductory part
presenting the Lagrangian formalism for relativistic fields and
Schwinger's
quantization  scheme, included the original axiomatic construction
of the scattering matrix based essentially on Bogoliubov's causality
principle, the renormalization method resting on the distribution
theory, as well as the functional integral method and the
generalized Tomonaga--Schwinger    equation. \par
       Technically, the book was written as follows. I would visit
N.N. in Lenin Hills and we would talk for an hour or two sketching the
next chapter. Then, in my place I would write the first version of
the text. At the next meeting this piece would be discussed and
frequently altered substantially. When approved, the
rewritten fair copy of the manuscript would be put in the left corner
on the top of a large wardrobe. It would be taken from there to be
typed by Evgeniya Aleksandrovna. Lightly embossed multicoloured
paper was used for  typing. Such paper, made in a factory  in Riga,
was purchased specially for our work. N.N. liked it very much.
Different sections of the manuscript had different colours: blue,
yellow, light green, violet...~. Three copies were typed at once.
I would collect the typed sections from the opposite, right corner
of the wardrobe to enter the formulae. \par
      The third copy of the coloured sections collected into
chapters was read critically by colleagues working at N.N.'s
department in the Steklov Institute. This reading provided the first
``grinding--in''. Two large articles in Uspekhi Fiz.
Nauk\footnote{Published in 1955\cite{uspechi1},\cite{uspechi2}.}
were intended to provide the second one. The text of the book\cite{book57}
which appeared in September 1957, was therefore, in principle,
quite well ``ironed out'' and, except for the last two chapters
containing new material on the renormalization group and
dispersion relations, constituted, so to say, the
``third approximation''. \par
     Looking back, equipped with my later experience as an author,
I would say that the monograph consisting of 30--odd printer's
sheets was created rather quickly. This was because from the
very beginning N.N. had a clear plan and later the
entire written text in his mind.

\subsection{The birth of Bogoliubov's renormalization group}

  In the spring of 1955 a small conference on ``Quantum Electrodynamics
and Elementary Particle Theory'' was organized in Moscow. It took
place at the Lebedev (Physical) Institute in the first half of April.
Among the participants there were several foreigners, including Hu Ning
and Gunnar K\"all\'en. My brief presentation touched upon the
consequences of finite Dyson transformations for renormalized Green
functions and matrix elements in quantum electrodynamics (QED). \par
	Landau's survey lecture ``Fundamental Problems in QFT'',
in which the ultraviolet (UV) behaviour in quantum field theory was
discussed, constituted the central event of the conference.
Not long before, the problem of short-distance behaviour in QED was
advanced substantially in a series of articles by Landau, Abrikosov,
and Khalatnikov. They succeeded in constructing a closed approximation
of the Schwinger--Dyson equations, which turned out to be compatible
both with renormalizability and gauge covariance. This so-called
``three--gamma'' approximation admitted an explicit solution in the
massless limit and, in modern language, it resulted in the
summation of the leading UV logarithms. \par
	   The most remarkable fact was that this solution turned out
to be self--contradictory from the physical point of view because it
contained a ``ghost  pole'' in the renormalized amplitude of the photon
propagator, the difficulty of ``zero physical charge''. \par
	     At that time our meetings with N.N. were regular and
intensive because we were busy preparing the final text of the book.
N.N. was very interested in the results of Landau's group and presented
me with the general problem of evaluating their reliability by
constructing, for example, the second approximation
(including {\it next-to-leading UV logarithms}) of the
Schwinger--Dyson equations to verify the stability of the UV
asymptotics and the existence of a ghost pole. \par
	    At that time I would sometimes meet Alesha Abrikosov,
whom I had known well since our student years. Shortly after the
conference at the Lebedev Institute, Alesha told me about
Gell--Mann and Low's article, which had just appeared. The same
physical problem was considered in this paper, but, as he put it,
it was complex to understand and they had not succeeded in combining
it with the results obtained by their group. \par
      I looked through the article and presented my teacher with
a brief report on the methods and results, which included some
general assertions on the scaling properties of charge distribution
at short distances and rather complex functional equations. \par
	   The scene that followed my report was quite impressive.
On the spot, N.N. announced that Gell--Mann and Low's approach was
correct and very important: it was a realization of the
{\sl normalization group (la groupe de normalisation)} discovered
a couple of years earlier by Stueckelberg and Petermann in the
course of discussing the structure of the finite arbitrariness in the
matrix elements arising upon removal of the divergences. This group
is an example of the continuous transformation groups studied by
Sophus Lie. This implied that functional group equations similar to
those obtained in the article by Gell--Mann and Low must take place
not only in the UV limit, but also in the general case. \par
	Then N.N. added that differential equations corresponding to
infinitesimal group transformations constitute the most powerful
tool in the theory of Lie groups. \par
	 Fortunately, I was also familiar with the foundations of
group theory. Within the next few days I succeeded in recasting
Dyson's finite transformations and obtaining the desired functional
equations for the scalar propagator amplitudes in QED, which have
group properties, as well as the corresponding differential
equations, that is, the Lie equations of the renormalization group.
Each of these resulting equations contained a specific object,
namely, the product of the squared electron charge  $\alpha=e^2$
and the transverse photon propagator amplitude   $d(Q^2)$.
We called this product, $e^2(Q^2)=e^2 d(Q^2)$, the {\sl invariant charge}.
>From the physical point of view it is an analogue of the so--called
{\sl effective charge} of an electron, first considered by Dirac
in 1933, which describes the effect of charge screening due to quantum
vacuum polarization. Also, the term ``renormalization group'' was first
introduced by us in the original publication \cite{bs-55a} in
Doklady Akademii Nauk SSSR in 1955 (and in Nuovo Cimento\cite{nc-56}
in 1956).


\subsection{Episode with a ''ghost pole''}
	    At the above--mentioned conference at the Lebedev Institute
Gunnar K\"all\'en presented a paper    written in collaboration with
Pauli on the so--called ''Lee model'', the exact solution of which
contained a {\sl ghost  pole} (which, in contrast to the physical one
corresponding to a bound state, had negative residue) in the nucleon
propagator. K\"all\'en and Pauli's analysis led to the conclusion
that the Lee model is physically void. \par
       In view of the result on the presence of a similar pole
in the photon propagator in QED (which follows from the solution of
Landau's group as well as an independent analysis by Fradkin)
obtained a little earlier in Moscow, K\"all\'en's report resulted in a
heated discussion on the possible inconsistency of QED. I remember
particularly well a scene by a blackboard on which K\"all\'en was
presenting an example of a series converging non-uniformly with
respect to a parameter (the terms of the series being dependent
on the parameter) to support the claim that no rigorous conclusion
about the properties of an infinite sum can be drawn from the analysis
of a finite number of terms. \par
	  The parties left without convincing one another and before
long a publication by Landau and Pomeranchuk appeared with a statement
that not only quantum electrodynamics, but also local
quantum field theory were self--contradictory. \par
	  Without going into details, let me remark that the
analysis of this problem carried out by N.N. with the aid of the
renormalization group formalism just developed by himself led to the
conclusion that such a claim cannot have the status of a
{\sl rigorous result, independent of perturbation theory}. \par
	 Nevertheless, like K\"all\'en's arguments, our work also
failed to convince the opponents. It is well known that Isaak
Yakovlevich Pomeranchuk even closed his quantum field theory
seminar shortly after these events.

\section{HISTORY OF~THE ~RENORMALIZATION ~GROUP ~IN ~QUANTUM
             ~FIELD ~THEORY}

\subsection{Renormalizations~ and~ re\-nor\-ma\-li\-za\-tion invariance}

     As is known, the regular formalism for eliminating ultraviolet
divergences in quantum field theory (QFT) was developed on the basis
of covariant perturbation theory in the late 40s. This breakthrough
is connected with the names of Tomonaga, Feynman, Schwinger and some
others. In particular, Dyson and Abdus Salam carried out the
general analysis of the structure of divergences in arbitrarily
high orders of perturbation theory. Nevertheless, a number
of subtle questions concerning so-called overlapping divergences
in the scattering matrix, as well as surface divergences, discovered by
Stueckelberg\cite{st-51}
in the Tomonaga--Schwinger equation, remained unclear. \par
	  An important contribution in this direction based on a
thorough analysis of the mathematical nature of UV divergences was
made by Bogoliubov. This was achieved on the basis of a branch of
mathematics which was new at that time, namely, the Sobolev--Schwartz
{\sl theory of distributions}. The point is that propagators
in local QFT are distributions (similar to the Dirac delta--function)
and their products appearing in the coefficients of the expansion
of the scattering matrix require supplementary definitions. In view
of this the UV divergence existence reflects the ambiguity in the
definition of the products in the case when their arguments coincide
or lie on the light cone. \par
        In the mid 50s on the basis of this approach Bogoliubov and
his disciples developed a technique of supplementing the definition
of the products of singular Stueckelberg--Feynman propagators
\cite{uspechi1}
and proved a theorem~\cite{paras1,paras2}    on the finiteness and
uniqueness (for renormalizable theories) of the scattering matrix
in any order of perturbation theory. The prescription part of this
theorem, namely, {\sl Bogoliubov's R-operation}, still remains a
practical means of obtaining finite and    unique results in
perturbative calculations in QFT. \par
	 The $R$--operation works, essentially, as follows. To remove
the UV divergences, instead of introducing some regularization, for
example, the momentum cutoff, and handling quasi-infinite counterterms,
it suffices to complete the definition of divergent Feynman integrals
by subtracting from them certain polynomials in the     external
momenta which in the simplest case are reduced to the first few
terms of the Taylor series of the integral. The uniqueness of
computational results is ensured by special conditions imposed on
them. These conditions contain specific degrees of
freedom\footnote{These degrees of freedom correspond to different
renormalization schemes \\ and  momentum scales.} that can be used to
establish the relationships between the Lagrangian parameters (masses,
coupling constants) and the corresponding physical quantities. The
fact that physical predictions are independent of the arbitrariness
in the renormalization conditions, that is, they are {\sl
renormalization--invariant}, constitutes the conceptual foundation
of the renormalization group. \par
        An attractive feature of this approach is that it is free
from any auxiliary nonphysical attributes such as bare masses,
coupling constants, and regularization parameters which turn out to
be unnecessary in computations employing Bogoliubov's approach.
As a whole, this method can be regarded as
{\sl renormalization without regularization and counterterms}.
\subsection{The discovery of the renormalization group}

	   The renormalization group approach has been known in
theoretical physics since the mid 50s. The renormalization group
was discovered by Stueckelberg and Petermann~\cite{stp} in 1953
as a group of infinitesimal transformations related to the
finite arbitrariness arising in the elements of the scattering
$S$-matrix upon elimination of the ultraviolet divergences. This
arbitrariness can be fixed by means of certain parameters $c_i$:\\
 \begin{quote}
{\sl
''... we must expect that a group of infinitesimal operators
 $\mbox{\bf P}_i=(\partial/\partial c_i)_{c=0}$,
exists, satisfying
$$
\mbox{\bf P}_iS=h_i(m,e)\partial S(m,e,...)/\partial e~,
$$
admitting thus a renormalization of $e$.''\\}
\end{quote}
\noindent
These authors introduced the {\sl normalization group} generated
(as a Lie group) by the infinitesimal operators $\mbox{\bf P}_i$
connected with the renormalization of the coupling constant $e$.
\par
    In the following year, on the basis of Dyson's transformations
written in the regularized form, Gell-Mann and Low~\cite{gml}
derived functional equations for QED propagators in the UV limit.
For example, for the renormalized transverse part $d$ of the
photon propagator they obtained an equation of the form
\begin{equation}
\label{1}
d\left(\frac{k^2}{\lambda^2},e_2^2\right)=
\frac{d_C(k^2/m^2,e_1^2)}{d_C(\lambda^2/m^2,e_1^2)}~,~~
e_2^2=e_1^2d_C(\lambda^2/m^2,e_1^2)~,
\end{equation}
where $\lambda$  is the cutoff momentum and  $e_2$ is the physical
electron charge. The appendix to this article contains the general
solution (obtained by T.D.Lee) of this functional equation for the
photon amplitude $d(x,e^2)$ written in two equivalent forms:
$$
e^2d\left(x,e^2\right)=F\left(xF^{-1}\left(e^2\right)\right)
$$
and
\begin{equation}
\label{2}
\ln x=\int\limits_{e^2}^{e^2d}\frac{{\rm d}y}{\psi(y)}~,
\end{equation}
where
$$
\psi(e^2)=\frac{\partial(e^2d)}{\partial\ln x}~~~\mbox{at}~~~x=1~. $$
A qualitative analysis of the behaviour of the electromagnetic
interaction at small distances was carried out with the aid of
(\ref{2}). Two possibilities, namely, infinite and finite charge
renormalizations were pointed out: \\

 \begin{quote}
{\sl
\noindent
Our conclusion is that the  {\bf shape} of the charge distribution
surrounding a test charge in the vacuum does not, at small distances,
depend on the coupling constant except through the scale factor. The
behavior of the propagator functions for large momenta is related
to the magnitude of the renormalization constants in the theory.
Thus it is shown that the unrenormalized coupling constant
$e_0^2/4\pi\hbar c$, which appears in perturbation theory as a power
series in the renormalized coupling constant $e_1^2/4\pi\hbar c$
with divergent coefficients, many behave either in two ways:\\

It may really be infinite as perturbation theory indicates;

It may be a finite number independent of
$e_1^2/4\pi\hbar c$. \\}
 \end{quote}

\noindent The latter possibility corresponds to the case when $\psi$
vanishes at a finite point:\footnote{Here  $\alpha_\infty $ is the
so--called fixed point of the renormalization group \\ transformations.}
$\psi(\alpha_\infty)=0$.

We remark that paper~\cite{gml} neither paid attention to the
group character of the analysis and the results obtained, nor
paper~\cite{stp} quoted. Moreover, the authors did not recognize
that the Dyson transformations used by them are valid only for the
transverse scaling of the electromagnetic field. Maybe this is why
they failed to establish a connection between their results and the
standard perturbation theory and they did not discuss the
possibility that a ghost pole might exist.

The final step was taken by Bogoliubov and Shirkov~\cite{bs-55a},
\cite{bs-55b} \footnote{
See also the survey~\cite{nc-56} published in English in 1956.}.
Using the group properties of finite Dyson transformations for the
coupling constant and the fields, the authors obtained functional
group equations for the propagators and vertices in QED in the
general case (that is, with mass taken into account). For example,
the equation for the transverse amplitude of the photon propagator
was obtained in the form
$$ d(x,y;e^2)=d(t,y;e^2) d\left(x/t, y/t; e^2d(t,y;e^2)\right)~,
$$
in which the dependence of $d$ not only on $x=k^2/\mu^2$ (where $\mu$
is a certain normalizing scale factor), but also on the mass variable
$y=m^2/\mu^2$ is taken into account.
\par
   In the modern notation, the above relation \footnote{In the
massless case $y=0$  it is equivalent to (\ref{4}).} is an equation
for the square of the effective electromagnetic coupling constant
$\bar\alpha =\alpha d(x,y;\alpha=e^2)$:
\begin{equation}
\label{3}      \bar\alpha(x,y;\alpha)=
\bar\alpha\left(x/t, y/t; \bar\alpha(t,y;\alpha)\right)~.
\end{equation}
The term {\sl ''renormalization group''} was introduced and
the notion of {\sl invariant charge} \footnote{This notion is now
 known as the effective or running coupling constant.}
was defined in ~\cite{bs-55a}.                                 

     Let us emphasize that, in contrast to the Gell--Mann and
Low approach, in our case there are no simplifications due to the
massless nature of the ultraviolet asymptotics. Here the homogeneity of
the mass scale is violated explicitly by the scale term $m$.
 Nevertheless, the symmetry (even though a bit more complex
one) underlying the renormalization group can, as before, be
stated as an {\sl exact symmetry} of the solutions of the
quantum field problem.\footnote{See equation  (\ref{10}) below.}.
This is what we mean when using the term {\sl Bogoliubov's
renormalization group} or {\bf Renorm-group} for short.

The following differential group equations for $\bar\alpha$:
\begin{equation}\label{4}
\frac{\partial\bar\alpha(x,y;\alpha)}{\partial\ln x}=
\beta\left(\frac{y}{x},\bar\alpha(x,y;\alpha)\right)
\end{equation}
in the nonlinear form, which is standard in Lie theory,
and for the electron propagator  $s(x,y;\alpha)$:
\begin{equation} \label{5}
\frac{\partial s(x,y;\alpha)}{\partial\ln x}=
\gamma\left(\frac{y}{x},\bar\alpha(x,y;\alpha)\right)s(x,y;\alpha)~,
\end{equation}
where
\begin{equation} \label{6}
\beta(y,\alpha)=\frac{\partial\bar\alpha(\xi,y;\alpha)}{\partial\xi}~,~
~~~\gamma(y,\alpha)=\frac{\partial s(\xi,y;\alpha)}{\partial\xi}~~~~
\mbox{at}~~\xi=1~.
\end{equation}
were first obtained by differentiating the functional equations.
In this way an explicit realization of the differential equations
mentioned in the citation from~\cite{stp}
was obtained. These results established a conceptual link between the
Stueckelberg--Petermann and Gell-Mann--Low approaches.


\subsection{Creation of the RG method}

Another important achievement of~\cite{bs-55a}
consisted in formulating a simple algorithm  for improving an
approximate perturbative solution by combining it with the
Lie equations\footnote{ Modern notation is used in this
quotation from~\cite{bs-55a} }:\\

\begin{quote}
{\sl
Formulae (\ref{4}) and (\ref{5}) show that to obtain expressions for
$\bar\alpha$ and $s$ valid for all values of their arguments one
has only to define \\ $\bar\alpha(\xi,y,\alpha)$ and $s(\xi,y,\alpha)$
in the vicinity of $\xi=1$. This can be done by means of the usual
perturbation theory.}
\end{quote}

In the next publication~\cite{bs-55b} this algorithm was used
effectively to analyse the ultraviolet and infrared (IR) asymptotic
behaviour in QED in transverse gauge. The one-loop and
two-loop UV asymptotics
\begin{equation} \label{7}
\bar\alpha^{(1)}_{RG}(x,0,\alpha)=
\frac{\alpha}{1-\frac{\alpha}{3\pi}\cdot\ln x}
\end{equation}
and
\begin{equation}   \label{8}
\bar\alpha^{(2)}_{RG}(x,0,\alpha)=
\frac{\alpha}{1-\frac{\alpha}{3\pi}\ln x
+\frac{3\alpha}{4\pi}\ln(1-\frac{\alpha}{3\pi}\ln x)}
\end{equation}
of the photon propagator as well as the IR asymptotics
$$
s(x,y;\alpha)\approx (p^2/m^2-1)^{-3\alpha/2\pi}~$$
of the electron propagator were obtained. At that time these
expressions had already been known only at the one--loop level. It
should be noted that in the mid 50s the problem of the UV behaviour
in local QFT was quite urgent. Substantial progress in the analysis
of QED at small distances was made by Landau and his
collaborators~\cite{landau} by solving an approximate version of the
Schwinger--Dyson equations including only the two-point functions
(``dressed" propagators) $\Delta_i(...,\alpha)$ and the three-point
function $\Gamma(...,\alpha)$, that is, the so-called ``three-gamma
equations''. The authors obtained asymptotic expressions for QED
propagators and 3-vertex, in which (using modern language) the leading
UV logarithms were summed \footnote{ These results were obtained under
arbitrary covariant  gauge.}. However, Landau's approach did not
provide a prescription for constructing subsequent approximations.

An answer to this question was given only within the new renormalization
group method. The simplest UV asymptotics of QED propagators obtained
in our paper~\cite{bs-55b}, for example, expression   (\ref{7}),
agreed precisely with the results of Landau's group.

Within the renormalization group approach these results can be
obtained in just a few lines of argument. To this end, the one-loop
approximation
$$
\bar\alpha^{(1)}_{PTh}(x,0;\alpha)=\alpha+
\frac{\alpha^2}{3\pi}\ell+...~~,~~~~\ell=\ln x $$
of perturbation theory should be substituted into the right-hand side
of the first equation in (\ref{6}) to compute the generator
$\beta(0,\alpha)=\psi(\alpha)= \alpha^2/3\pi$, followed by an
elementary integration.

Moreover, starting from the two-loop expression
$$
\bar\alpha^{(2)}_{PTh}(x,0;\alpha)=\alpha+
\frac{\alpha^2}{3\pi}\ell+\frac{\alpha^2}{\pi^2}\left(\frac{\ell^2}{9}
+\frac{\ell}{4}\right)+...~,  $$
we arrive at the second renormalization group approximation (\ref{8})
corresponding to the summation of the next-to-leading UV logarithms.
This demonstrates that the RG method is a regular procedure, within
which it is quite easy to estimate the range
of applicability of the results.

The second--order renormalization group solution (\ref{8}) for the
effective coupling constant first obtained in~\cite{bs-55b} contains
the nontrivial log--of--log dependence which is now widely known
as the two--loop approximation for the running coupling constant in
quantum chromodynamics (QCD).

Before long~\cite{sh-55} this approach was formulated for the case
of QFT with two coupling constants $g$ and $h$, namely, for a model
of pion--nucleon interactions with self-interaction of pions \footnote{
It is essential that for the Yukawa PS $\pi N$--interaction  $\sim g$
to be renorma- \\ lizable, it is necessary to add to the Lagrangian a
quartic pion self-interaction \\
term with an independent,  that is, a second, coupling constant $h$.
At that time \\ this was not fully recognized: compare  the given
system with equations \\ (4.19)$^\prime$-(4.21)$^\prime$ in~\cite{gml},
and the discussion in \cite{dau-bohr}.}.
The following system of two coupled equation:
\begin{eqnarray}
&&\bar g^2\left(x,y;g^2,h\right)=
\bar g^2\left(\frac x t, \frac y t, \bar g^2(t,y; g^2, h),
\bar h\left(t,y;g^2,h\right)\right)~,\nonumber\\
&&\bar h\left(x,y;g^2,h\right)=
\bar h\left(\frac x t, \frac y t, \bar g^2\left(t,y;g^2,h\right),
\bar h\left(t,y;g^2,h\right)\right)~.\nonumber
\end{eqnarray}
was first obtained. The corresponding system of nonlinear
differential equations from \cite{sh-55} was used in~\cite{ilya}
to carry out the UV analysis of the renormalizable model of
pion-nucleon interactions based on one-loop perturbative computations.

In~\cite{bs-55a,bs-55b} and  ~\cite{sh-55} the renormalization group
approach was thus directly connected with practical computations
of the UV and IR asymptotics. Since then this technique, known as
the {\sl renormalization group method} (RGM), has become the sole
means of asymptotic analysis in local QFT.

\subsection{Other early RG applications}

The first general theoretical application of the RG method was made
in the summer of 1955 in connection with the (then topical)
so-called ghost pole problem (also known as the ``zero-charge trouble").
This effect, first discovered in QED~\cite{zero1,zero2},
was at first thought~\cite{zero2} to indicate a possible
difficulty in quantum electrodynamics, and then \cite{dau-bohr,zero3}
as a proof of the inconsistency of the whole local QFT.

However, the renormalization group analysis of the problem carried
out in~\cite{bsh56}  on the basis of (\ref{2}) demonstrated that no
conclusion obtained with the aid of finite--order computations
within perturbation theory can be regarded as a complete proof.
This corresponds precisely to the impression, one can get when
comparing (\ref{7}) and (\ref{8}). In the mid 50s this result was very
significant, for it restored the reputation of local QFT. Nevertheless,
in the course of the following decade the applicability of QFT in
elementary particle physics remained doubtful in the eyes of many
theoreticians.

In the general case of arbitrary covariant gauge the renormalization
group analysis in QED was carried out in~\cite{tolia}. Here the problem
is that the charge renormalization is connected only with the
transverse part of the photon propagator. Therefore, under nontransverse
(for example, Feynman) gauge the Dyson transformation has a more
complex form. Logunov proposed to solve this problem by considering the
gauge parameter is another coupling constant.

Ovsyannikov~\cite{oves} found the general solution of the functional
RG equations taking mass into account:
$$
\Phi(y,\alpha)=\Phi\left(y/x, \bar\alpha(x,y;\alpha)\right)~$$
\noindent in terms of   an arbitrary function $\Phi$
of two arguments, reversible in its second argument. To solve
the equations, he used the differential group equations
represented as linear partial differential equations of the
form\footnote{Which are now known as the Callan---Symanzik
equations.}:
$$
\left\{x\frac{\partial}{\partial x}+y\frac{\partial}{\partial y}
-\beta(y,\alpha)\frac{\partial}{\partial\alpha}\right\}
\bar\alpha(x,y,\alpha)=0~.
$$

The results of this ``period of pioneers'' were  collected
in the chapter ``Renormalization group'' in the monograph ~\cite{kniga},
the first edition of which appeared in 1957
\footnote{ Shortly after that is was translated into English
and French~\cite{book}.}, and very quickly acquired the status of
``quantum--field folklore''.

\section{FURTHER RG DEVELOPMENT}

\subsection{Quantum field theory}

The next decade brought a calm period, during which there was
practically no substantial progress in the renormalization group
method. An important exception, which ought to be mentioned here, was
Weinberg's article~\cite{steve}, in which the idea of a {\sl running
mass} of a fermion was proposed. If considered from the viewpoint
of~\cite{tolia}, this idea can be formulated as follows:
\begin{quote}
{\sl any parameter of the Lagrangian can be treated as a (generalized)
coupling constant, and so can be included into the renormalization
group formalism}.
\end{quote}

However, the results obtained in the framework
of this approach turned out the same as before. For example,
the most familiar expression for the fermion running mass
$$  \bar m(x,\alpha)=m_\mu\left(\frac{\alpha}{\bar\alpha(x,
\alpha)}\right)^\nu~,
$$
in which the leading UV logarithms are  summed, was known for the
electron mass in QED (with  $\nu=9/4$) since the mid 50s
(see~\cite{landau} и \cite{bs-55b}).

New possibilities for applying the RG method were discovered when the
technique of operator expansion at small distances (on the light cone)
appeared. The idea of this approach stems from the fact that the RG
transform, regarded as a Dyson transformation of the renormalized
vertex function, involves the simultaneous scaling of all its invariant
arguments (normally, the squares of the momenta) of this function.
The expansion on the light cone, so to say, ''separates the
arguments'', as a result of which it becomes possible to study
the physical UV asymptotic behaviour by means of the expansion
coefficients (when some momenta are fixed and lie on mass shell).
The argument--separation method for functions of several variables
proposed by Wilson makes it possible to study a number of cases
important from the physical point of view.

The end of the calm period can be marked well enough by the year
1971, when the renormalization group method was applied in the
quantum theory of non-Abelian gauge fields, in which the famous
effect of asymptotic freedom was discovered~\cite{12}.

The renormalization group expression
$$
\bar\alpha_s^{(1)}=\frac{\alpha_s}{1+\beta_1\ln x}~,       $$
for the effective coupling constants $\bar\alpha_s$ in QCD, computed
in the one-loop approximation, exhibits a remarkable UV asymptotic
behaviour thanks to $\beta_1$ being positive. This expression implies,
in contrast to Eq.~(\ref{7}), that the effective QCD constant decreases
as $x$ increases and tends to zero in the UV limit. This discovery,
which has become technically possible only because of the RG method use,
is the most important physical result obtained with the aid of the RG
approach in particle physics.

\subsection{Spin lattice}

At the same time Wilson~\cite{13} succeeded in transplanting the
RG philosophy from relativistic QFT to another branch of modern
theoretical physics, namely, the theory of phase transitions in spin
systems on a lattice. This new version of the RG was based on
Kadanoff's idea\cite{leo} of joininig in "blocks" of few neighbouring
spins with appropriate change (renormalization) of the coupling
constant.

To realize this idea, it is necessary to average the spins in each
block. This operation reducing the number of degrees of freedom and
simplifying the system under consideration, preserves all its
long-distance properties under a suitable renormalization of the
coupling constant. Along with this, the above procedure gives rise to
a new theoretical model of the original physical system.

In order that the system obtained by averaging be similar to the
original one, one must also discard some terms of the new effective
Hamiltonian which turns out to be unimportant in the description of
infrared
properties. As a result of this {\sl Kadanoff -- Wilson decimation},
we arrive at a new model system characterized by new values of the
elementary scale and coupling constant. By iterating this operation,
one can construct a discrete ordered set of models. From the physical
point of view the passage from one model to some other one
is an irreversible approximate operation. Two passages of that sort
applied in sequence are equivalent to one, which gives rise to a group
structure in the set of models. However, in this case the renormalization
group is an approximative and is realized as a semigroup.

This construction, obviously in no way connected with UV properties, was
much clearer from the general physical point of view and could therefore
be readily understood by many theoreticians. Because of this, in the
seventies the concept of the renormalization group and its algorithmic
structure were rather quickly and successfully carried over to
new branches of theoretical physics such as polymer physics~\cite{14},
the  theory of noncoherent transfer~\cite{15}, and so on.

Apart from constructions analogous to those of Kadanoff and
Wilson, in a number of cases the connection with the original
quantum field RG was established.

\subsection{Turbulence}

This has been done with help of the functional integral representation.
For example, the classic Kolmogorov--type turbulence problem was
connected with the RG approach by the following steps~\cite{16}:
\begin{enumerate}
\item Define the generating functional for correlation functions.
\item Write for this functional the path integral representation.
\item By a change of functional integration variable establish an
     equivalence of the given classical statistical system with some
     QFT model.
\item Construct the Schwinger--Dyson equations for this equivalent QFT.
\item Apply the Feynman diagram technique and perform the finite
      renormalization procedure.
\item Write down the standard RG equations and use them to find fixed
      point and scaling behavior.
\end{enumerate}
The physics of renormalization transformation in the turbulence problem
is related to a change of UV cutoff in the wave-number variable.

\subsection{Ways of the RG expanding}

As we can see, in different branches of physics  the renormalization
group developed in two directions:
\begin{itemize}
\item
 The construction of a set of models for the physical problem at
hand by direct analogy with the Kadanoff -- Wilson construction
(averaging over certain degrees of freedom) --- in polymer physics,
noncoherent transfer theory, percolation theory, and others;
\item   The search for an exact RG symmetry in the theory directly or
by proving its equivalence to some QFT: for example, in turbulence
theory ~\cite{16,sasha}, turbulence in plasma~\cite{17},
phase transition physics (based on a model of a
continuous spin field).
\end{itemize}

What is the nature of the symmetry underlying the
renormalization group?

а) In QFT the RG symmetry is an exact symmetry of a solution described
in terms of the fundamental notions of the theory and some boundary
value(s).

b) In turbulence  and some other branches of physics it is a symmetry
  of a solution of an equivalent quantum field model.

c) In spin lattice theory, polymer theory, noncoherent transfer theory,
percolation theory, and so on (in which the blocking concept of
Kadanoff and Wilson is applied) the RG transformation involves
transitions between various auxiliary models (constructed especially
for this purpose). To formulate RG, it is necessary to construct an
ordered set ${\cal M}$ of models $M_i$. The RG transformation
connecting various models has the form
$$
R(n)M_i=M_{ni}~. $$
In this case the RG symmetry can thus be realized only
on the whole set  ${\cal M}$.

There is also a purely mathematical difference between the aforesaid
realizations of the renormalization group. In field theory the RG
is a continuous symmetry group. On the contrary, in the
theory of critical phenomena, polymers, and other similar cases
(when an averaging operation is necessary) we have an
approximate discrete semigroup. It must be pointed out that in
dynamical chaos theory, in which renormalization group ideas and
terminology can sometimes be applied too, functional
iterations do not constitute a group at all, in general.
An entirely different terminology is    sometimes adopted
in the above--mentioned domains of theoretical physics.
Expressions such as ''the real--space renormalization group'',
''the Wilson RG'', ''the Monte--Carlo RG'', or
''the chaos RG'' are used.

Nevertheless, the affirmative answer to the question

\centerline{\underline{``Are there distinct renormalization groups?''}}
\noindent implies no more than what has just been said about the
differences
between cases a) and b) on the one hand and c) on the other.

\subsection{Two faces of the renormalization group in QFT}

As has been mentioned above, invariance under RG transformations,
that is, renorm-group invariance, is a very
important notion in renormalized quantum field   theory. It
means that all physical results are independent of the
choice of the renormalization scheme and the subtraction point.
The latter corresponds to a symmetry whose presence is
embodied in the renormalization group. In QFT the RG
transformations can be considered in two different ways.

The existence of virtual states and virtual processes is a
characteristic feature of quantum field theory. For example,
virtual transformations of a photon into an electron--positron
pair and vice versa can take place in QED. This vacuum polarization
process gives rise to the notion of effective charge. In the classical
theory of electromagnetism a test electric charge placed in a
polarizable medium attracts nearby charges of opposite sign, which
leads to partial screening of the test charge. In QED the vacuum,
that is, the very space between the particles, serves as a
polarizable medium. The electron charge is screened by the vacuum
fluctuations of the electromagnetic field. Dirac was the first
to demonstrate in 1933~\cite{dirac} that the electron charge in
momentum representation depends on $Q^2$ according to the formula
\begin{equation}\label{8a}
e(Q^2,\Lambda^2)=e_0\left\{1+\frac{\alpha_0}{6\pi}\ln\frac{Q^2}
{\Lambda^2}+...\right\}~,
\end{equation}
where $e_0=\sqrt{\alpha_0}$ is the bare charge and  $\Lambda$
is the cutoff momentum.

The first attempt to formulate renormalization group ideas in this
context was undertaken by Stueckelberg and Petermann~\cite{stp}.
In their pioneering work the RG transformations were introduced
somewhat formally being related to the procedure for divergences
eliminating, the result of which contains a finite arbitrariness.
It is this ``degree of freedom'' in the finite renormalized
expressions that was used in our papers~\cite{bs-55a},~\cite{bs-55b}.
Roughly speaking, our result corresponds to a parameter change
$(\Lambda\to\mu)$ describing the degree of freedom, so that
the ''finite representation''
\begin{equation}\label{9}
e(Q^2,\mu^2)=e_\mu\left\{1+\frac{\alpha_\mu}{6\pi}
\ln\frac{Q^2}{\mu^2} +...\right\}~,
\end{equation}
is obtained in place of Eq.~(\ref{8a}), $e_\mu=\sqrt{\alpha_\mu}$
being the physical charge of an electron measured at $Q^2=\mu^2$.
Here the renormalization group symmetry can therefore be expressed in
terms of the momentum transfer scale, that is renormalization point $\mu$.

Gell-Mann and Low used another representation. In their article the
small distance behaviour in QED is analysed in terms of $\Lambda$, the
momentum transfer cutoff. In this approach the electron charge
could be represented by the expression (\ref{8a}) that is singular
in the limit $\Lambda \to \infty$.

We shall present a simple physical picture (which can be derived
from Wilson's Nobel lecture) to illustrate this approach. Imagine
an electron of finite dimensions  distributed in a small volume of
radius  $R_\Lambda=\hbar/c\Lambda$ with $\ln(\Lambda^2/m_e^2)\gg1$.
We assume that the electric charge of such a nonlocal electron
depends on the cutoff momentum so that this dependence accumulates
the effects of vacuum polarization taking place at distances not
exceeding  $R_\Lambda$ from the centre of the electron.
We thus obtain a set of models with a nonlocal electron of charges
$e_i=\sqrt{\alpha_i}$ corresponding to different values of the
cutoff parameter  $\Lambda_i$.

Here $\alpha_i$ depends on $R_i$ as the effects of vacuum polarization
in the excluded volume  $r<R_i$ must be subtracted. In this picture
the RG transformation can be thought of as the passage from one
radius $R_i=\hbar/c\Lambda_i$ to another $R_j$ accompanied
by a simultaneous change of the effective electron charge
$$
e_i\equiv e(\Lambda_i)\to e_j=e_i\left(1+\frac{\alpha_i}{6\pi}
\ln\frac{\Lambda_j^2}{\Lambda_i^2}+...\right),
$$
which can be determined with the aid of  (\ref{8a}). In other words,
here the RG plays the role of a symmetry of operations in the
space of nonlocal QED models constructed so that each model is
equivalent to the true local theory at long distances. It is
right to say that in these two approaches the renormalization
groups differ from one another.

\subsection{Functional self--similarity}

An attempt to analyse the relationship between these formulations
on a simple common basis was undertaken about ten years ago~\cite{sh-82}.
In this paper (see also our surveys \cite{sh-84,O-7,KEK}) it was
demonstrated that all the above--mentioned realizations of the
renormalization group could be considered in a unified manner by using
only some common notions of mathematical physics.

In the general case it proves convenient to discuss the symmetry
underlying the renormalization group with the aid of a continuous
one--parameter transformation of two variables $x$ and $g$ written as
\begin{equation}   \label{10}
R_t:\left\{x\to x^\prime=x/t,~g\to g^\prime=\bar g(t,g)\right\}~.
\end{equation}
Here $x$ is the basic variable subject to a scaling transformation,
while $g$ is a physical quantity undergoing a more complicated
functional transformation. To form a group, the transform $R_t$ must
satisfy the multiplication law
$$
R_tR_\tau=R_{t\tau}~,
$$
which leads to the following functional equation for $\bar g$:
\begin{equation}\label{12}
\bar g(x,g)=\bar g\left(x/t, \bar g(t,g)\right)~.
\end{equation}
This equation has the same form as the functional equation (\ref{3})
for the effective coupling in QFT in the massless case, that is,
when  $y=0$. It is also fully equivalent to the Gell-Mann--Low
functional equation (\ref{1}). It is therefore clear that the
contents of RG equation can easily be reduced to the
group multiplication law.

In physical problems the second argument $g$ of the transformation is
usually the boundary value of a dynamical function, that is, a
solution of the problem under investigation. This means that the
symmetry underlying the RG approach is a symmetry of the solution
(not of the equation) describing the physical system at hand,
involving a transformation of the parameters entering the boundary
conditions.

As an illustration, we consider a solution
$f(x)$ defined by the boundary condition  $f(x_0)=f_0$.
Among the arguments of $f$ we also include the boundary
parameters: $f(x)=f(x,x_0,f_0)$. In this case the RG transformation
corresponds to altering the parametrization of the solution, say,
from  $\{x_0,f_0\}$ to $\{x_1,f_1\}$. In other words, the value of
$x$  for which the boundary condition is given should not be
equal to  $x_0$ (that is, another point  $x_i$ can also be used).
We now assume that  $f$ can be represented as   $F(x/x_0,f_0)$
with  $F(1,\gamma)=\gamma$. The equality
$$
F\left(x/x_0, f_0\right)=F\left(x/x_i, f_i\right)
$$
reflects the fact that the function itself is not modified under
that change of the boundary condition\footnote{
As, for example, in the case $F(x,\gamma)=\Phi(\ln x+\gamma).$}.
Setting $f_1=F(x_1/x_0,f_0)$, $\xi=x/x_0$ and $t=x_1/x_0$,
we get $F(\xi,f_0)=F(\xi/t,F(t,f_0))$,
which is equivalent to  (\ref{12}). The group operation can now
be defined by analogy with Eq.~(\ref{10}):
$$
R_t~:~\{~\xi\to \xi/t~,~~f_0\to f_1=F(t,f_0)~\}~. $$
Therefore, in the simplest case the RG can be defined as a continuous
one--parameter group of transformations of a solution of the
physical problem fixed by a boundary condition. The RG transformation
affects the parameters of the boundary condition and corresponds to
changing the way in which this condition is introduced for
{\sl one and the same solution}.

Special cases of such transformations have been known for a long time.
If we assume that $F= \bar g$ is a factored function of its arguments,
then from Eq.(\ref{12}) it follows that $F(z,f)=fz^k$, with $k$ being
a number. In this particular case the group transform takes the form
$$
P_t~:~\{~z\to z^\prime=z/t~,~~f\to f^\prime=ft^k~\}~,
$$
that is known in mathematical physics as a power {\sl self-similarity
transformation}. More general case $R_t$ with functional transformation
law can be characterized as a {\sl functional self--similarity}
(FSS) transformation~\cite{sh-82}.

\section{C O N C L U S I O N}

We can now answer the question concerning the physical meaning
of the symmetry that underlies functional self-similarity and the
renormalization group. Consider the case when the RG is equivalent to
FSS. As we have already mentioned, it is not a symmetry of the
physical system or the equations of the problem at hand, but a
symmetry of a solution considered as a function of the essential
physical variables and suitable boundary conditions. A symmetry like
that can be defined, in particular, as the invariance of a physical
quantity described by this solution with respect to the way in which
the boundary conditions are inposed. Changing this way constitutes a
group operation in the sense that the group property can be considered
as the transitivity property of such changes.

Homogeneity is an important feature of the physical systems under
consideration. However, homogeneity can be violated in a discrete manner.
Let us imagine that such a discrete inhomogeneity is connected with
a certain value of $x$, say, $x=y$. In this case the RG transformation
with canonical parameter $t$ will have the form:
$$
R_t~:~\{~x^\prime=x/t~,~~y^\prime=y/t~,~~g^\prime=\bar g(t,y;g)~\}~.
$$
The group multiplication law yields precisely the functional
equation~(\ref{3}).

The symmetry connected with functional self--similarity is a very
simple and frequently encountered property of physical phenomena.
It can easily be ``discovered'' in many very different problems of
theoretical phy\-sics: in classical mechanics, transfer theory,
classical hydrodynamics, and so on ~\cite{mamikon,O-7,KEK,RG91}.

 Recently, interesting attempts have been made~\cite{КP-а,venia}
to use the RG concept in classical mathematical physics, in particular,
to solve nonlinear differential equations. These articles discuss the
possibility of establishing a regular method for finding a special
class of symmetries of the equations in modern mathematical physics,
namely, RG-type symmetries. The latter are defined as solution
symmetries with respect to transformations involving parameters that
enter into the solutions through the equations as well as through the
boundary conditions in addition to (or even rather than) the natural
variables of the problem present in the equations (see \cite{19,KKPrg}).

As is well known, the aim of modern group analysis \cite{Oves,Ibr},
which goes back to works of S.~Lie\cite{Lie}, is to find symmetries
of differential equations (DE). This approach does not
include a similar problem of studying the symmetries of solutions of
these equations. Beside the main direction of both the classical and
modern analysis, there also remains the study of solution symmetries
with respect to transformations involving not only the variables
present in the equations, but also parameters appearing in the
solutions, including the boundary conditions.

>From the aforesaid it is clear that the symmetries which attracted
attention in the 50s in connection with the discovery of the RG in
QFT were those involving the parameters of the system in the group
transformations. It is natural to refer to these symmetries related
to functional self-similarity as the {\it RG-type symmetries}. As we
have already mentioned, they are inherent in many problems of
mathematical and theoretical physics. It is therefore important to
establish, on the basis of modern group analysis, a regular method
for finding RG symmetries for various classes of mathematical problems,
including those whose formulation goes beyond the scope of systems of
a finite number of partial differential equations.

The timeliness of the search for RG symmetries is connected with the
effectiveness of the RG method, which makes it possible to improve
the properties of approximate solutions of problems possessing the
symmetry and, in particular, to reconstruct the correct
structure of the behaviour of the solution in a neighbourhood of a
singularity, which is, as a rule, disturbed by the approximation.

In problems admitting description in terms of DE's a regular algorithm
for finding RG-type symmetries can be constructed by combining the
group analysis \cite{KKPrg,KKP94} with Ambartsumyan invariant embedding
method \cite{Ambar}. In those cases when the embedding of the
boundary--value problem for a DE leads to an integral formulation,
it is required that the algorithms of group analysis should be
extended to integro--differential systems of equations.
Taking into account that recently some progress has also
been made~\cite{in-dif1,in-dif2,in-dif3} in extending the range of
applicability
of the established methods of group analysis, one can say that the
above combination turns out to be constructive enough also for
integro--differential equations. We recall that the first
embedding with a physical end in view was realized for the integral
equation of radiative transfer~\cite{Ambar}.

At the same time, the embedding of the Cauchy problem for systems of
ordinary differential equations brings us back to the origins of the
theory of such equations. This is because it can be realized within
the framework of the    well--known theorem   on the existence of
derivatives with respect to the initial values of the solutions of the
system. Here it proves fruitful to treat the parameters (such as a
coupling constant) as new variables introduced into the group
transformations and/or the embedding procedure.

The differential formulation of RG symmetries employs an
infinitesimal operator (tangent vector field), which, in general,
combines the symmetry of the original problem with a symmetry
(approximate or exact) of its solution taking the boundary
conditions into account. Algorithmically, the invariant embedding
procedure contains the operation of including these data among
the variables taking part in group transformations. Here the object
of group analysis is the system of equations consisting of the
initial system and the embedding equations corresponding to it
and to the boundary--value problem. The latter can be constructed
on the basis of both the original equations and the boundary
conditions. From the viewpoint of group  analysis, combining the
original system with the embedding equations changes the differential
manifold (as a rule, quite substantially).

          The symmetry group of the combined system can be found by
the usual methods of Lie analysis and its modern modifications with
the aid of the solution of the determining equation for the
coordinates of the infinitesimal operator corresponding to the
condition that ensures the invariance of this new
manifold. As a matter of fact, the RG can be obtained
(see Refs.[57,58]) by a suitable restriction of the resulting
group to a solution, the representation of which can be quite diverse:
as an exact integral or an algebraic expression, as a final
portion of a perturbation series or another approximation formula,
as a functional integral, and so on.

The author would like to express his gratitude to Drs. B.V.~Medvedev
and V.V.~Pustovalov for helpful remarks.


\vspace{5mm}
\addcontentsline{toc}{section}{References}

\begin{thebibliography}{99}
\bibitem{nn-50/51}  N.N.~Bogoliubov,  {\sl Dokl.Akad.Nauk SSSR}, {\bf 81}
              (1951), 757-760, 1015-1018.
\bibitem{uspechi1} N.N.~Bogoliubov and D.V.~Shirkov, "Problems in quantum
   field theory. I", {\sl Uspekhi Fiz. Nauk} {\bf 55}, (1955), 149-214.
\bibitem{uspechi2} N.N.~Bogoliubov and D.V.~Shirkov, "Problems in
   quantum field theory.II", {\it ibid.} {\bf 57}, (1955), 3-92.
\bibitem{book57} N.N.~Bogoliubov and D.V.~Shirkov, {\sl Introduction to the
        theory of quantized fields}, Gostekhizdat, Moscow,  1957.
	English and French translations: see \cite{book}         
\bibitem{bs-55a} N.N.~Bogoliubov and D.V.~Shirkov, "On the renormalization
    group  in quantum electrodynamics", {\sl Doklady AN SSSR},
        {\bf 103} (1955) 203-206.                                     
\bibitem{bs-55b} N.N.~Bogoliubov and D.V.~Shirkov, "Application of the
        renormalization group to improve the formulae of perturbation
   theory", {\sl Dokl.Akad. Nauk SSSR}, {\bf 103} (1955) 391-394.
\bibitem{nc-56}  N.N.~Bogoliubov  and D.V.~Shirkov,~ "Charge       
	renormalization group in quantum  field theory",
	{\it Nuovo Cim.} {\bf 3} (1956) 845-637.
\bibitem{st-51} E.C.G.~Stueckelberg and J.~Green, {\sl Helv. Phys.
            Acta},   {\bf 24} (1951) 153-174.
\bibitem{paras1} N.N.~Bogoliubov and  O.S.~Parasyuk, {\sl Dokl.Akad.
      Nauk SSSR},   {\bf 100} 	(1955) 25--28, 429--432;
\bibitem{paras2} N.N.~Bogoliubov  and O.S.~Parasyuk,~ {\it Acta             
           Mathematica},  {\bf 97} (1957), 227--266.
\bibitem{stp} E.C.G.~Stueckelberg  and A.~Petermann, "La normalisation
       des constantes dans la theorie des quanta".                   
       {\sl Helv. Phys. Acta}, {\bf 26} (1953) 499-520.
\bibitem{gml} M.~Gell-Mann  and F.~Low, "Quantum Electrodynamics at
      Small Distances", {\sl Phys. Rev.} {\bf 95} (1954) 1300-1312.
\bibitem{landau}  L.D.~Landau, A.A.~Abrikosov and I.M.~Khalatnikov,         
        {\sl Doklady  AN SSSR}, {\bf 95} (1954) 497-499; 773-776;
    1117-1120; {\bf 96} (1954) 261-264; {\sl Nuovo Cim.} Supp.3, 80-104.
\bibitem{sh-55} D.V.~Shirkov, "Renormalization group with two coupling
    constants in the theory of pseudoscalar mesons",
      {\sl Dokl.Akad. Nauk SSSR}, {\bf 105} (1955) 972-975.
\bibitem{ilya} I.F.~Ginzburg,  {\it ibid.}, {\sl Dokl.Akad. Nauk SSSR}
            {\bf 110} (1956) 535--538.
\bibitem{zero1} E.S.~Fradkin, {\sl Zh. Eksp. Teor. Fiz.} {\bf 28}
            (1955) 750-752;
English transl. in {\sl Soviet Phys. JETP} {\bf 1} (1955).
\bibitem{zero2}  L.D.~Landau and I.Ya.~Pomeranchuk, {\sl                      
                Dokl.Akad. Nauk SSSR}, {\bf 102}  (1955) 489-492.
\bibitem{dau-bohr}  L.D.~Landau, "On the Quantum Theory of Fields" In: {\it Niels Bohr and the development of physics},
      eds. W.~Pauli et al., Pergamon, London, 1955, pp 52-69.
\bibitem{zero3} I.Ya.~Pomeranchuk, {\sl Dokl.Akad. Nauk SSSR}, {\bf 103} (1955) 1005--1008;
              {\bf 105} (1955) 461--464; {\sl Nuovo Cim.} {\bf 10} (1956) 1186-1203.
\bibitem{bsh56} N.N.~Bogoliubov and D.V.~Shirkov,~ "Lie type model in
       quantum electrodynamics", {\sl Dokl.Akad. Nauk SSSR} {\bf 105} (1955) 685-688.
\bibitem{tolia} A.A.~Logunov, {\sl Zh. Eksp. Teor. Fiz.} {\bf 30} (1956) 793-795; English transl. in Soviet Phys. JETP {\bf 3} (1956).      
\bibitem{oves} L.V.~Ovsyannikov, {\sl Dokl. Akad. Nauk SSSR}, {\bf 109}
             (1956) 1112-1115.
\bibitem{kniga}  N.N.~Bogoliubov and D.V.~Shirkov, {\sl Introduction to the
   theory of quantized fields}, Nauka, Moscow 1957, 1973, 1976 and 1984.
   English and French translations: see \cite{book}.
\bibitem{book} N.N.~Bogoliubov  and D.V.~Shirkov, {\sl Introduction to
   the theory of quantized fields};  Two American editions 1959 and 1980,
      Wiley-Interscience, N.Y.;
  Bogolioubov N.N., Chirkov D.V.~ {\sl Introduction \`a la
      Theorie des Champes Quantique}, Paris, Dunod, 1960.               
\bibitem{steve} S.~Weinberg, {\sl Phys.Rev.}, {\bf D8} (1973) 605-625.
\bibitem{conus} K.~Wilson, {\it Phys.Rev.} {\bf 179} (1969) 1499-1515.
\bibitem{12} D.~Gross and P.~Wilczek,~ {\sl Phys.Rev.}, {\bf D8} (1973)
 3633-3652; \par H.~Politzer.~ {\sl Phys.Rev.Lett.} {\bf 30} (1973) 1346-1349.
\bibitem{13} K.~Wilson,~ "RG and Critical Phenomena",
         {\sl Phys.Rev.}, {\bf B4} (1971) 3174-3183.                    
\bibitem{leo} L.~Kadanoff,~ {\it Physica} {\bf 2} (1966) 263.
\bibitem{14} P.G.~De Gennes,~ {\sl Phys.Lett.}, {\bf 38A} (1972) 339-340;\\
       J.~des Cloiseaux,~ {\sl J.Physique (Paris)}, {\bf 36} (1975) 281.
\bibitem{15} T.L.~Bell et al.~ "RG Approach to Noncoherent Radiative
           Transfer", {\sl Phys.Rev.}, {\bf A17} (1978) 1049-1057;         
          G.F.~Chapline,~ {\sl Phys.Rev.}, {\bf A21} (1980) 1263-1271.
\bibitem{16} C.~DeDominicis, {\sl Phys.Rev.}, {\bf A19} (1979) 419-422.
\bibitem{sasha} L.~Adjmyan, A.~Vasil'ev, and M.~Gnatich,~                     
      {\sl Teoret. Mat. Fiz.}, {\bf 58} (1984) 72; {\bf 65} (1985) 196; see also  \\
       A.N.~Vasiliev,~ "Quantum Field Renormalization Group in the
        Theory of Turbulence and in Magnetic Hydrodynamics"  in
        \cite{rg-86} , pp 146-159.
\bibitem{rg-86} {\sl Renormalization Group}, (Proc. 1986 Dubna Conference),
	   Eds. D.V.~Shirkov, D.I.~Kazakov, and A.A.~Vladimirov,
      World Scientific, Singapore, 1988.
\bibitem{17} G.~Pelletier,~ {\sl Plasma Phys.}, {\bf 24} (1980) 421.
\bibitem{chir} B.V.~Chirikov and D.L.~Shepelansky, "Chaos Border and
     Statistical Anomalies",  pp 221-250 --  in \cite{rg-86}.
\bibitem{dirac} P.A.M.~Dirac~ in {\sl Theorie du Positron} (7-eme
         Conseil du Physique Solvay: Structure et propriete de noyaux
          atomiques, Octobre 1933), Gauthier-Villars, Paris,1934,
          pp.203-230.
\bibitem{sh-82} D.V.~Shirkov,~ "The renormalization group, the invariance
     principle, and functional self-similarity", {\sl Dokl.Akad. Nauk SSSR},
      {\bf 263} (1982) 64-67; English transl. in {\sl Soviet Phys. Dokl.}
      {\bf 27} (1982).
\bibitem{sh-84} D.V.~Shirkov,~ in {\sl Nonlinear and turbulent processes
        in physics}, Ed. R.Z.~Sagdeev, Harwood Acad.Publ., N.Y. 1984,
        v.{\bf 3}, pp~1637-1647.
\bibitem{O-7} D.V.~Shirkov,~"Renormalization group in modern physics",
             in \cite{rg-86}  pp 1-32; also {\sl Int. J. Mod.Phys.}      
              {\bf A3} (1988), 1321-1341.
\bibitem{KEK} D.V.~Shirkov,~ "Renormalization group in different
      fields of theoretical physics", KEK Report 91-13, Feb. 1992. (English).
\bibitem{mamikon} M.A.~Mnatsakanyan.~{\sl Dokl.Akad. Nauk SSSR}, {\bf 262} (1982) 856-860.
       English transl. in {\sl Soviet Phys. Dokl.} {\bf 27} (1982).
\bibitem{RG91} D.V.~Shirkov,~ "Several topics on renorm-group theory".
           In \cite{rg-91} , pp~1-10.
\bibitem{rg-91} {\sl Renormalization Group '91}. (Proceedings of 1991 Dubna
            Conf.), Eds. D.V.~Shirkov, V.B.~Priezzhev, World
            Scientific, Singapore, 1992.
\bibitem{КP-а} V.F.~Kovalev and V.V.~Pustovalov,~"Strong nonlinearity and
          generation of higher harmonics in nonuniform plasma",
          Preprint No 78, Lebedev Fiz.Inst. AN SSSR, 1987.
\bibitem{venia} V.F.~Kovalev and V.V.~Pustovalov,~ "Functional
  self-similarity in a problem in the theory of plasma with electron
  nonlinearity", {\sl Teoret. Mat. Fiz.} {\bf 81} (1989), No 1(10), 69-85.
\bibitem{19} N.~Goldenfeld, O.~Martin  and Y.~Oono,~ "Intermediate
           asymptotics and renormalization group theory",
           {\sl J.Sci.Comput.(USA)}, {\bf 4} (1990), 355-72.
\bibitem{KKPrg} V.F.~Kovalev, S.V.~Krivenko and V.V.~Pustovalov,~
    "The RG method based on group analysis", in \cite{rg-91} pp.300-314.
\bibitem{Oves} L.V.~Ovsyannikov,~{\sl Group analysis of differential
         equations}, Nauka, Moscow  1978 (Russian) and Academic Press,
	 N.Y., 1982.
\bibitem{Ibr} N.Kh.~Ibragimov,~{\sl Transformation group in mathematical
        physics}, Nauka, Moscow 1983; English transl. {\sl Transformation
	groups applied to mathematical physics}, Reidel,
	Dordrecht--Lancaster 1985.
\bibitem{Lie} Sophus Lie.~ {\it Gesammelte Abhandlungen.} Leipzig--Oslo,
            Bd.5, 1924; Bd 6, 1927.
\bibitem{KKP94} V.F.~Kovalev, S.V.~Krivenko, and V.V.~Pustovalov,
     "Lie symmetry and group for the solution of a boundary-value problem".
    {\sl Differentsial'nye Uravneniya}, {\bf 30} No 10, (1994).
	  English transl. in {\sl Differential Equations} {\bf 30} (1994).
\bibitem{Ambar} V.A.~Ambartsumyan,~{\sl Astr. Zh.} {\bf 19} No 5,
      (1942), 30-41; {\sl Dokl.Akad. Nauk SSSR} {\bf 38} No 8, 257 (1943);
     {\sl Izv. Akad. Nauk Arm. SSR}, Estestv. Nauki,  No 1-2, p. 37 (Russian).
\bibitem{in-dif1} Yu.N.~Grigor'ev and S.V.~Meleshko,~"Group analysis
     of the integro-differential Boltzmann equation", {\sl Dokl.
   Akad. Nauk SSSR} {\bf 297} (1987), 323-327; English transl. in
		{\sl Soviet Phys. Dokl.} {\bf 32} (1987).
\bibitem{in-dif2} V.I.~Fushchich, V.M.~Shtelen' and N.I.~Serov,~{\sl
  Symmetry analysis and exact solutions of nonlinear equations in
  mathematical physics} Naukova Dumka, Kiev  1989  (Russian).
\bibitem{in-dif3} V.F.~Kovalev, S.V.~Krivenko, and V.V.~Pustovalov.~
    "Group analysis of Vlasov's kinetic eq.", {\sl Differentsial'nye
    Uravneniya}, {\bf 29} (1993) 1804-1817; 1971-1983; English transl.
    in  {\sl Differential Equations} {\bf 29} (1993).
\bibitem{lga} V.F.~Kovalev, S.V.~Krivenko, V.V.~Pustovalov. ~"Lie
  algebra of renorma\- lization group admitted by initial-value
  problem for Burgers equation", {\it Lie Groups and their Aplications
  (Istanbul)},  v.{\bf 1}, No 2, (1994).
\bibitem{in-dif} V.V.~Pustovalov and A.K.~Shvarev, "Group and
renormalization group analysis of the equations of nonlinear geometrical
 optics. 1. Lie symmetries", {\sl Preprint, Lebedev Institute }., No 15,
 1994, 87 с.

\end{thebibliography}
\end{document}